\documentclass[aps,pre,floatfix,twocolumn,nofootinbib]{revtex4}
\usepackage{bm}
\usepackage[dvips]{epsfig}
\usepackage{subfigure}
\usepackage{graphicx} 
\usepackage{amsmath} 
\usepackage{amssymb}

\usepackage{xcolor}

\def\bea{\begin{eqnarray}}
\def\eea{\end{eqnarray}}

\def\bro{\boldsymbol{\rho}}

\newcommand{\comment}[1]{}
\newcommand{\BEQ}{\begin{align}}
\newcommand{\EEQ}{\end{align}}
\newcommand{\BEA}{\begin{eqnarray}}
\newcommand{\EEA}{\end{eqnarray}}
\newcommand{\nn}{\nonumber \\}

\renewcommand{\d}{{\rm d}}

\newcommand{\bk}{{\bf k}}

\newcommand{\bq}{\boldsymbol{q}}

\newcommand{\br}{{\bf r}}

\renewcommand{\bro}{\boldsymbol{\rho}}

\newcommand{\rar}{\rightarrow}
\newcommand{\qE}{\mathsf{E}}

\begin{document}

\title{Quantum Aberrations: Entangling Photons with Zernike
Polynomials
}
\author{H. Avetisyan}
\affiliation{
Alikahanyan National Laboratory (Yerevan Physics Institute), 2 Alikhanyan
Brothers Street, Yerevan 0036, Armenia}
\email{h.avetisyan@yerphi.am}
\author{G. Nikoghosyan}
\affiliation{Yerevan State University, 1 Alex Manoogian Street, Yerevan, 0025, Armenia}

\begin{abstract}
We introduce Zernike polynomials as a novel degree of freedom for encoding quantum information in the spatial structure of photons. Building on their orthogonality and completeness over the unit disc, we develop a framework for generating, manipulating, and detecting photons in Zernike modes, and propose methods for realizing single-photon and two-photon Zernike wave packets. We demonstrate analytically that two-photon states generated via spontaneous parametric down-conversion exhibit mode entanglement in the Zernike basis, with correlations arising from selection rules enforced by Clebsch–Gordan coefficients. Our results open a new pathway for structured spatial entanglement, complementary to schemes based on Laguerre-Gaussian or Hermite-Gaussian modes, and suggest practical experimental implementations based on holographic modulation and optical Fourier techniques.
\end{abstract}
\maketitle
 
\section{Introduction}

Structured spatial modes of light have emerged as a powerful resource for quantum information processing. In particular, encoding quantum states in orbital angular momentum or Hermite-Gaussian spatial modes has led to demonstrations of high-dimensional entanglement, quantum communication protocols, and quantum information processing \cite{Allen, Bazhenov, Beijersbergen, Padgett, Agarwal, Mair, Vaziri}. 
However, extending the set of spatial mode bases beyond conventional families such as Hermite-Gaussian and Laguerre-Gaussian opens new possibilities for control, robustness, and efficiency.

Historically, a great deal of attention has been paid to the Zernike modes as a complete set of polynomials to expand the aberration phase on the exit pupil of an imaging optical system. We direct the reader to Zernike's groundbreaking 1934 paper \cite{Zernike}, Nijboer's 1942 thesis \cite{nijboer} on the diffraction theory of aberrations, and to Born \& Wolf \cite{B&W}, particularly Chapter 9, Sections 9.1–9.4, and Appendix VII, for an exploration of the early development of the Nijboer-Zernike theory and a comprehensive review of it, including the mathematical properties of the polynomials. 
Here we mention only that Zernike disc polynomials $Z(x,y)$ are distinguished from other sets of polynomials by their elegant invariance properties: when the coordinate axes are rotated about the origin, each polynomial is transformed into a polynomial of the same form \cite{bhatia-wolf} in the sense that $Z(x,y)=G(\theta)Z(x',y')$ ($x',y'$ are the rotated coordinates). More importantly, each Zernike polynomial represents an optimally balanced combination of Seidel power series terms \cite{nijboer}, minimizing variance across the pupil – or equivalently, maximizing intensity in Gaussian focus \cite{mahajan}. Furthermore, Zernike polynomials are mathematically related to generalized spherical harmonics, which arise in the solution of Laplace’s equation in higher-dimensional spaces \cite{Zernike_Brinkman}.
It should be noted that the true origin of the ad hoc operator introduced by Zernike has not yet been adequately addressed. A recent attempt has been made in \cite{Vahagn}.

Relatively little work has been carried out in the quantum domain using Zernike modes or the Zernike Hamiltonian \cite{poghosyan}, leaving significant opportunities for further exploration and development in this area.
For example, Zernike modes can be used to represent qudits. For example, Zernike modes $Z_n^m$ can be used to represent qudits, with the mode indices $n$ and $m$ serving as quantum numbers. These indices can be manipulated—raised or lowered—using optical masks generated through interferometric techniques. \cite{Heckenberg}. One can also measure the Zernike modes using the same masks to convert the Zernike modes into plane-wave modes. In quantum information schemes, first-order Zernike modes can be manipulated in a manner analogous to the linear and circular polarization of electromagnetic fields \cite{oneil}. Devices can be constructed to act on these modes in a way analogous to the operation of polarizing beam splitters, half-wave plates, and quarter-wave plates \cite{Xue, sasada}. As a result, first-order Zernike modes can form a basis for encoding qubits and enabling single-qubit rotations. We outline methods for constructing photon states and defining field operators in the Zernike basis. Moreover, we show that photon pairs entangled in Zernike modes can be generated through the process of spontaneous parametric down-conversion (SPDC). Spatial mode entanglement of down-converted photons has already been demonstrated, both theoretically and experimentally, for Laguerre-Gaussian and Hermite-Gaussian modes \cite{walborn04, walborn05}.

Although Hermite-Gaussian and Laguerre-Gaussian modes are widely used in quantum optics and are compatible with circularly symmetric systems, they are not specifically tailored to represent optical aberrations. In contrast, Zernike modes form a complete orthogonal basis on the unit disc and directly correspond to common wavefront aberrations, making them naturally suited for describing and manipulating spatial distortions in pupil-limited systems. Zernike modes are ideally suited for quantum protocols involving optical imperfections, adaptive optics, or pupil-limited imaging systems.

In this work, we present the first complete quantum framework for the use of Zernike modes in quantum optics. Our new results include:

\noindent
(i) A formal construction of photon creation and annihilation operators in the Zernike basis;

\noindent
(ii) Analytical expressions for single-photon and two-photon wave packets in this basis;

\noindent
(iii) Demonstration of two-photon entanglement via spontaneous parametric down-conversion (SPDC), with selection rules governed by Clebsch–Gordan coefficients;

\noindent
(iv) Experimental strategies for preparing, transforming, and detecting Zernike modes; and

\noindent
(v) Discussion of specific quantum tasks, such as aberration sensing and robust communication, where Zernike entanglement offers distinct advantages.

These results extend the scope of spatial-mode-based quantum optics by introducing a new, physically grounded basis that is especially well matched to practical optical systems exhibiting pupil-limited or aberrated behavior.


\section{Zernike polynomials and some of their properties}\label{Zernik}
Zernike polynomials are not only a convenient mathematical basis; they are rooted in the physics of optical aberrations. In classical optics, they provide a complete and orthogonal basis over the unit disc, allowing for compact representation of wavefront distortions across a circular pupil. Each low-order Zernike polynomial corresponds to a distinct aberration type commonly encountered in imaging systems. For instance, $Z_2^0$ represents defocus, $Z_3^{\pm 1}$ corresponds to coma, and $Z_2^{\pm 2}$ describes astigmatism. Higher-order terms represent more complex distortions such as spherical aberration or trefoil. These modes have been extensively used in optical testing, adaptive optics, and phase correction systems, where the aberration phase $\Phi(\rho,\theta)$ is expanded in the Zernike basis and manipulated directly using deformable mirrors or phase masks. In interferometry, the Zernike coefficients are extracted to quantify system imperfections, while in vision science they are used to characterize human eye aberrations \cite{malacara, dai}. 

By leveraging this strong physical foundation, we propose (see Sec. \ref{Z1phot}) extending the use of Zernike modes into quantum optics, where they can serve as spatial degrees of freedom for encoding quantum information. Their well-defined aberration structure and rotational symmetries make them physically meaningful and experimentally accessible, rather than purely abstract mathematical constructs.

\begin{figure}[h!]
    \centering
    \includegraphics[width=0.8\linewidth]{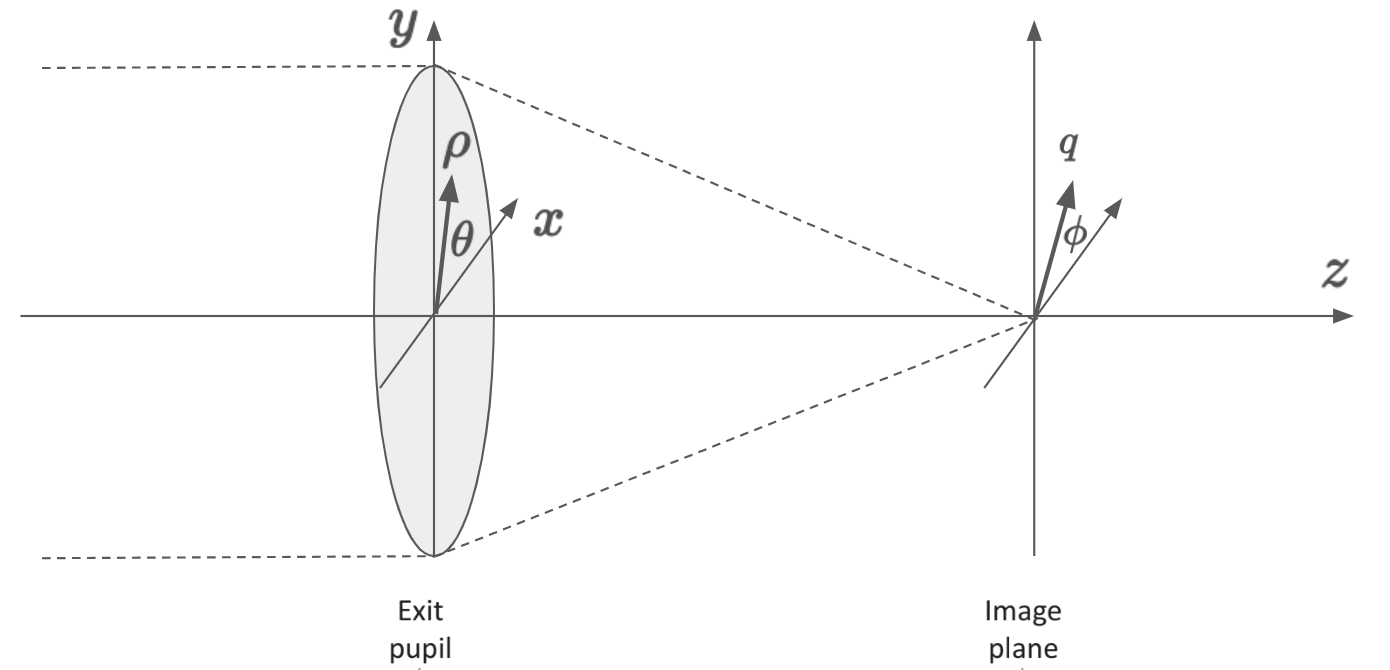}
    \caption{The basic optical configuration. A wavefront in the exit pupil coming from a point in the object plane (not shown) converges towards the image plane. The position on the exit pupil is defined by the polar coordinates $(\rho, \theta)$; the position in the image plane region is defined by the polar coordinate system $(q, \phi)$.}
    \label{fig:pupil}
\end{figure}

We work with generalized pupil functions $P$ characterized by a non-uniform amplitude transmission function $A$ and a real phase aberration $\Phi$. We adopt the definition of Zernike disc polynomials using exponential rather than trigonometric azimuthal dependence. Accordingly, the generalized pupil function 
$P$ is expanded, e.g., in polar coordinates, $\rho = \sqrt{x^2+y^2}, \quad \theta = \arctan(y/x)$, see Fig.\ref{fig:pupil}, as 
\begin{align}
    P(\rho, \theta) &= A(\rho, \theta) \exp[i\Phi(\rho, \theta)] 
    \nn&=
    \sum_{n=0}^\infty \sum_{m=-n}^n a_{mn} Z_n^m(\rho, \theta),\nn
    &0 \leq  \rho \leq 1,\qquad
    0 \leq  \theta < 2\pi, \label{pupil_exp}
\end{align}
in which the normalized polynomials are expressed in terms of radial polynomials $R_n^{|m|}(\rho)$ \cite{B&W} as
\begin{align}
Z_n^m(\rho,\theta) = \sqrt{n+1}\, R_n^{|m|}(\rho) e^{im\theta}\label{Zern},
\end{align}
\begin{align}
R_n^{|m|}(\rho,\theta) = \sum_{k=0}^{\frac{n-|m|}{2}}\frac{(-1)^k(n-k)!}{k!\left(\frac{n+m}{2}-k\right)!\left(\frac{n-m}{2}-k\right)!}\rho^{n-2k},
\end{align}
and where $n$ and $m$ in the summation are integers such that $n - |m|$ is even and non-negative. The coefficients $a_{mn}$ are given by 
\begin{align}
    a_{mn} = \frac{1}{\pi} \int\d^2\rho\ 
    P(\bro)Z_n^m(\bro),
\end{align}
where $\bro = (\rho,\theta)$, $\d^2 \rho = \rho\,\d\rho\d\theta$, and the integration is performed over the pupil region \eqref{pupil_exp}.
The form \eqref{Zern} provides great simplification in computing diffraction integrals due to the separation of variables.
When the system is aberration-free, $A=1, \Phi = 0$ in the pupil. Note that most authors in the optics community consider uniform amplitude transmission pupils, $A=1$, and expand only the aberration phase $\Phi$ into Zernike polynomials. 

The orthogonality and completeness relations of the polynomials are as follows
\begin{align}
    \int\d^2\rho\,Z_n^{m*}(\bro) Z_{n'}^{m'}(\bro) &= \pi\delta_{nn'}\delta_{mm'}\label{orthoZ},\\
    \sum_{mn}Z_n^{m*}(\bro) Z_{n}^{m}(\bro') &= \pi\delta(\bro-\bro').\label{completeness}
\end{align}
The number of linearly independent polynomials of degree $\leq n$ is $\frac{1}{2}(n+1)(n+2)$ \cite{mahajan}. Hence, every polynomial in $(x,y)$ can be expressed as a linear combination of a finite number of polynomials $Z_n^m$, which means the set is complete \cite{B&W}.
A product of two polynomials are linearized as \cite{tango, vanhaver}
\begin{align}
    Z_{n_1}^{m_1}(\bro)Z_{n_2}^{m_2}(\bro) = \sum_{n_3} A_{n_1n_2n_3}^{m_1m_2,m_1+m_2} \, Z_{n_3}^{m_1+m_2}(\bro).\label{zprod}
\end{align}
Here the summation is over all non-negative integers $n_3$ with 
\begin{align}
    \text{parity}(n_3)=\text{parity}(m_3)=\text{parity}(m_1+m_2),\label{CG0}
\end{align}
$m_3=m_1+m_2$ such that $\max(|m_3|,|n_1-n_2|)\leq n_3\leq n_1+n_2.$
There are at most $\min\{n_1,n_2\} + 1$ terms in the sum.
The $A$'s are given by the Clebsch-Gordan coefficients \cite{varshal}
\begin{align}
    A_{n_1n_2n_3}^{m_1m_2m_3}= \sqrt{\frac{n_3+1}{(n_1+1)(n_2+1)}}\left|C_{\frac{m_1}{2}\frac{m_2}{2}\frac{m_3}{2}}^{\frac{n_1}{2}\frac{n_2}{2}\frac{n_3}{2}}\right|^2,
\end{align}
which are non-vanishing only when 
\begin{align}
n_1, n_2, n_3 = \text{integers or half-integers }\geq 0,
\end{align}
such that  
\begin{align}
\frac{n_1}{2}+\frac{n_2}{2}+\frac{n_3}{2}= \text{ integer}
\end{align}
while satisfying the triangle conditions 
\begin{align}
|n_r - n_s| \leq n_t \leq n_r + n_s
\end{align}
for any permutation $r,s, t$ of $1, 2, 3,$ and when $m_r = -n_r, -n_r + 1, \ldots, n_r - 1, n_r, \quad r = 1, 2, 3,$
with $m_1 + m_2 - m_3 = 0.$

The Fourier transforms of the Zernike polynomials are computed to be \cite{Noll}
\begin{align} 
    \Tilde{Z}_n^m(\bq)
    &=\int_0^1 \d \rho\,\rho\int_0^{2\pi}
    \d\theta\,Z_n^m(\rho,\theta)e^{2\pi i \rho q \cos(\theta-\phi)}
   \nn&=
   2\pi i^n\sqrt{n+1}\frac{J_{n+1}(2\pi q)}{2\pi q} e^{im\phi},
  \label{Z-fourier}
  \end{align}
where $\bq = (q,\phi)$.
Note that the index $m$ is associated only with the angular factor $e^{im\phi}$ while the index $n$ — with radial factor $J_{n+1}(2\pi q)/(2\pi q)$.
The $\Tilde{Z}_n^m$ are orthogonal and complete over the entire space,
\begin{align}\label{ZtildOrth}
    \int_{-\infty}^\infty \d^2q \,\Tilde{Z}_n^{m*}(\bq) \Tilde{Z}_{n'}^{m'}(\bq) &= \pi \delta_{mm'}\delta_{nn'},\\
    \sum_{mn}\Tilde{Z}_n^{m*}(\bq) \Tilde{Z}_{n}^{m}(\bq') &= \pi\delta(\bq-\bq').
\end{align}
as can be checked using \eqref{orthoZ},\eqref{completeness}, the orthogonality of sines and cosines as well as Bessel functions
\begin{align}
    \int_0^\infty \frac{\d t}{t}\, J_{m+2p+1}(t)J_{m+2q+1}(t) = \frac{\delta_{pq}}{2(m+2p+1)}.
\end{align}

\section{Transformations in Zernike basis}\label{Zdetection}
To utilize Zernike modes as a resource for quantum information processing, it is essential to establish viable methods for their experimental generation, manipulation, and measurement. In this section, we outline a sequence of practical procedures that enable full control over photon wave packets in the Zernike basis. We begin by describing methods for mode preparation using interferometric and holographic techniques. We then examine how these modes evolve under free-space propagation and optical transformation. Following this, we discuss mode conversion strategies based on the symmetry properties of Zernike modes, and finally, we present detection schemes that project incoming quantum states onto specific Zernike modes for measurement.
\subsection{Preparation}
The preparation of photons in specific Zernike modes is a crucial first step toward implementing quantum information protocols based on this basis. Zernike modes, defined on the unit disc, are suited to be encoded in the spatial profile of a light field across a circular aperture. 
To imprint these profiles onto photons, one can use spatial light modulation techniques that are already standard in classical optics and wavefront control.

One widely used method is the application of computer-generated holographic masks \cite{Fukushima, Heckenberg}. These are diffractive optical elements designed to transform an incoming Gaussian or plane wave into a desired, Zernike, mode in the present case, by modulating its amplitude and phase. Such holograms can be fabricated using spatial light modulators (SLMs), digital micromirror devices \cite{goorden}, and they can be dynamically reprogrammed to generate different modes on demand.

Another practical technique involves an interferometric generation of Zernike modes using a Twyman–Green interferometer \cite{mahajan}. In this setup, the beam is split into two parts and one arm is modified by inserting a known optical aberration (such as a lens with known aberration, or a deformable mirror). When the two arms are recombined, the resulting interference pattern encodes the desired Zernike phase structure.

For single-photon states, these classical field modulation methods can be directly applied to the spatial profile of a single photon using a transmissive or reflective SLM. The hologram effectively acts as a projection operator, transforming an incoming photon wavefront into a specific Zernike spatial profile. The success of this transformation can be verified by observing the resulting diffraction pattern in the Fourier plane or by mode-selective detection downstream.

Thus, the preparation of Zernike modes leverages existing tools in classical beam shaping and wavefront engineering, making this quantum-optical implementation both conceptually accessible and experimentally feasible.

    

\noindent
\subsection{Image space}\label{imgspc}
In Fourier optics, the image of an object is the convolution of the image predicted by geometrical optics with the Fraunhofer diffraction pattern of the exit pupil \cite{goodman}. Now, the Fourier transforms \eqref{Z-fourier} can be generated as Frounhofer pattern of the Zernike field profile. This can be done in the focal plane of a spherical lens with focal length $f$:
\begin{align}
    \Tilde{Z}_n^m(\bro) &= \frac{k}{2\pi f}\int\d\bro' \, Z_n^m(\bro')e^{-i\frac{k}{f}\bro\cdot \bro'}
    \nn&=
    2\pi i^n\sqrt{n+1} 
    \frac{J_{n+1}(2\pi\rho k/f)}{2\pi \rho} e^{im\phi}. \label{ZFraunh} 
\end{align}
The amplitude and phase of the light at coordinates $(\rho_x, \rho_y)$ in the focal plane are determined by the amplitude and phase of the input Fourier component at spatial frequencies $(q_x=k\rho_x/f, q_y=k\rho_y/f)$. From \eqref{ZFraunh} it is seen that when a pupil function is expanded in the Zernke basis as in \eqref{pupil_exp}, the spatial spectrum of the pupil function itself may be associated with the amplitude distribution in the image plane 
\begin{align}
    \Tilde{P}(\bq) = \sum_{mn} a_{mn} \Tilde{Z}_n^m(\bq),\label{pupil_spectr}
\end{align}
with the same expansion coefficients. This means that a function in the pupil plane expressed as a series of Zernike polynomials $Z_n^m(\bro)$ immediately yields its image plane function by the simple substitution of the $\Tilde{Z}_n^m(\bq)$ for $Z_n^m(\bro)$.
It should be noted that the series expansion for the \textit{wavefront} distortion in the pupil does not have this simple representation. This procedure may be used to calculate the diffraction patterns in the image plane of an optical system. 

\noindent
\subsection{Propagation}
Applying the Fresnel diffraction operator to the Zernike profile at $z=0$ one gets
\begin{align}
    Z_n^m(\bro,z)&= -\frac{ike^{ikz}}{z\pi}
    \int \d^2 \rho' Z_{n}^{m}(\bro') e^{\frac{ik}{2z}|\bro-\bro'|^2}
    \nn&= 
    -\frac{ik}{z}e^{ikz+\frac{ik\rho^2}{2z}}e^{im\theta}V_n^m(\rho;z) \label{ZFresnel}
\end{align}
where we integrated the theta part and defined
\begin{align}
    V_n^m(\rho;z) &\equiv i^{m}\sqrt{n+1}
    \int_0^1 \d \rho' \rho' R_{n}^{m}(\rho')
    \nn&\quad\times
    J_m\left(-\frac{k\rho\rho'}{z}\right) \exp\left[\frac{ik}{2z}\rho'^2\right]. \label{exp-coef-1pZ}
\end{align}
This integral has been evaluated by \cite{vanhaver} as Bessel-Bessel series. The coefficients $V_n^m$ are
\begin{align}
    V_n^m(\rho;z)&=
    \sqrt{n+1} 
    e^{-\frac{ik}{4z}}
    \sum_{h}\sum_{l=\frac{|n-h|}{2}}^{\frac{n+h}{2}} i^{l-h} (2l+1) 
    \nn&\times
    A_{2l,n,h}^{0,m,m}\,\, j_l\left(-\frac{k}{4z}\right)\frac{J_{h+1}(2\pi \rho)}{2\pi \rho},\label{vanhaver_exp}
\end{align}
where $j_l$ is the spherical Bessel function, and the summation is over all $h$
of same parity as $m$ with $h \geq |m|$.
Due to the decay properties of (spherical) Bessel functions with increasing order, the
number of terms to be used in the double series in \eqref{vanhaver_exp} should be slightly larger than $\pi\rho k/z$. The summations can be limited to values of $h$ up to a little beyond $\pi e\rho$ and values of $l$ up to a little beyond $e k/(8z)$ \cite{vanhaver}. Note also, that for Fraunhofer zone $z\gg k \times (\text{pupil area})$, only the spherical Bessel function for $l=0$ (and $h=n$) case $j_0(\beta)=\sin(\beta)/\beta\rightarrow 1$ survives the sum for small $\beta$, i.e., 
$V_n^m(\rho;z)\rightarrow \sqrt{n+1}\, i^{-n}\frac{J_{n+1}(2\pi \rho)}{2\pi\rho}$
and \eqref{ZFresnel} becomes proportional to \eqref{ZFraunh}.

\subsection{Mode conversion}
Due to the symmetry properties of Zernike modes, the modes with the same $n$-index and $\pm m$ differ from each other by rotation. Hence, these modes can be achieved one from the other by placing a beam rotator in between. The beam rotator is made of two dove prisms, and if one of them rotated about the optical axis by $\theta/2$, then the amplitude distribution of the field will rotate by $\theta$. To change the mode index $n$, conventional techniques can be used to add/subtract aberrations \cite{malacara}.

\subsection{Detection} To detect a photon in the Zernike basis, the holographic mask of the corresponding mode interferogram should be placed before the conventional detector system in the optical Fourier recognition setup \cite{goodman, malacara}. 

\section{Quantization in Zernike modes}\label{Z1phot}

In this section we proceed with the quantization procedure using Zernike modes. We define the Zernike mode creation operators as
\begin{align}\label{zmodop}
    \Tilde{z}_n^{m\dagger} &= \frac{1}{\sqrt{\pi}}\int \d^2q \, \Tilde{Z}_n^m(\bq) a^\dagger(\bq),\\
    a^\dagger(\bq) &= \frac{1}{\sqrt{\pi}}\sum_{mn}\Tilde{Z}_n^{m*}(\bq) \Tilde{z}_n^{m\dagger} ,
\end{align}
obeying the commutation relations
\begin{align}
    [\Tilde{z}_n^m, \Tilde{z}_{n'}^{m'\dagger}]&=
    \int \d^2q \, 
    \Tilde{Z}_n^{m*}(\bq)\Tilde{Z}_{n'}^{m'}(\bq)=\delta_{mm'}\delta_{nn'}
\end{align}
due to the usual relation $[a(\bq), a^\dagger(\bq')] = \delta(\bq-\bq')$. While the polarization degree of freedom could, in principle, be included in the analysis, it plays no essential role in the discussions that follow. For simplicity, we assume the creation operator is implicitly summed over polarization states, i.e., $a^\dagger(\bq) = \frac{1}{sq
2}\sum_s a^\dagger_s(\bq)$.
Then, it can be seen that the states defined as 
\begin{align}
    \Tilde{z}_n^{m\dagger} |0\rangle =|\Tilde{z}_n^m\rangle,\label{zstate}
\end{align}
are orthonormal and complete, 
\begin{align}
    \langle \Tilde{z}_n^m|\Tilde{z}_{n'}^{m'}\rangle&=\delta_{mm'}\delta_{nn'},\\
    \sum_{n=0}^\infty \sum_{m=-n(2)}^n |\Tilde{z}_{n}^{m}\rangle \langle \Tilde{z}_n^m|&=I,
\end{align}
where $\sum_{m=-n(2)}^n$ indicates that the sum is over $m$ in the range $\{-n,-n+2,\cdots, n\}$.
\subsection{The field operator and the first-order correlation function}
The angular spectrum representation \cite{B&W} of the field operator 
\begin{align}
\qE^{(+)}(\br) = 
    \int \frac{\d^2 q}{(2\pi)^2} \, a(\bq)
    e^{i \left(\bq\cdot\bro+\sqrt{k^2-q^2}z\right)}
\end{align}
in paraxial approximation ($\sqrt{k^2-q^2} \rar k - \frac{q^2}{2k}$) in terms of the Zernike mode annihilation operators $\Tilde{z}_{n}^{m}$ has the form
\begin{align}
    \qE^{(+)}(\br) &= \frac{e^{ikz}}{\sqrt{\pi}}\sum_{mn} \Tilde{z}_{n}^{m}
    \int \frac{\d^2 q}{(2\pi)^2} \, \Tilde{Z}_{n}^{m}(\bq)
    e^{i\left( \bq\cdot\bro-\frac{q^2z}{2k} \right)}
    \nn&=
    \frac{ke^{ikz}}{2iz\pi^\frac{3}{2}}\sum_{mn} \Tilde{z}_{n}^{m} 
    \int \d^2 \rho' Z_{n}^{m}(\bro') e^{\frac{ik}{2z}|\bro-\bro'|^2}
    \nn&=
    \frac{ke^{ikz+\frac{ik\rho^2}{2z}}}{iz\sqrt{\pi}} 
    e^{im\theta} \sum_{mn}V_n^m(\rho;z)\,\Tilde{z}_{n}^{m},
    \label{field-op-1pZ}    
\end{align}
where $V_n^m$ are given in \eqref{vanhaver_exp}.
The field operator assumes a particularly simple form in the Fraunhofer region (c.f. Sec. \ref{imgspc})
\begin{align}
    \qE^{(+)}(\br) &=e^{ikz}\sum_{mn}\Tilde{Z}_n^m(\rho_x,\rho_y)\,\Tilde{z}_{n}^{m},
    \label{field-op-1pZFrhf}
\end{align}
with $\rho_{x(y)}=fq_{x(y)}/k$.

\subsection{Single-photon state in the Zernike basis}
Now using the orthonormality of the Fourier transforms of the Zernike polynomials \eqref{ZtildOrth}, one can expand any state in the Zernike basis using their closure relation. Particularly, a paraxial single-photon state, $|\psi_1\rangle=\int\frac{\d^2 q}{(2\pi)^2}\, C (\bk)a^\dagger(\bk)|0\rangle$, where $C(\bk)$ is strongly concentrated
around $\bk_0 = k_0\hat{\textbf{z}}$ with a small transverse part $|\bq| \leq \vartheta_0k_0$, can be written as
\begin{align}
    |\psi_1\rangle &= \sum_{mn}  |\Tilde{z}_n^m\rangle \langle \Tilde{z}_n^m|\psi_1\rangle\equiv
    \sum_{mn} \zeta^m_n |\Tilde{z}_n^m\rangle,
\end{align}
where (relabeling $\bk$ using $\bq$)
\begin{align}
    \zeta^m_n &= \langle z_n^m|\psi_1\rangle
    =\int \d^2q \, C(\bq)\Tilde{Z}_n^{m*}(\bq).
\end{align}
From now on we will work only with paraxial states.
\subsection{First-order correlation function}
The first-order correlation function $G^{(1)}(\br)=\langle\psi_1|\qE^{(-)}(\br)\qE^{(+)}(\br)|\psi_1\rangle$, too,  assumes a particularly simple form in the Fraunhofer region
\begin{align}
    G^{(1)}(\br)= 
    \sum_{\substack{m_1m_2\\n_1n_2}}
    \zeta^{m_1}_{n_1}\zeta^{m_2*}_{n_2}
    \Tilde{Z}_{n_1}^{m_1}(\rho_x,\rho_y)\, \Tilde{Z}_{n_2}^{m_2*}(\rho_x,\rho_y).
    \label{G2}
\end{align}
Eq. \eqref{G2} satisfies the first-order coherence condition, as expected for any pure single-photon state.

\subsection{Two-photon state in the Zernike basis}\label{Z2phot}
The most general paraxial two-photon state 
\begin{align}
    |\psi_2\rangle =&\frac{1}{\sqrt{2}}  \iint \d^2q_1\d^2q_2 \, 
    C(\bq_1,\bq_2) a^\dagger(\bq_1) a^\dagger(\bq_2) |0\rangle, \label{2p}
\end{align}
\begin{align}
    \langle\psi_2 |\psi_2\rangle&=1, \label{normalization2}
\end{align}
can be represented in the two-photon Zernike basis. The latter can be constructed using tensor products of one-photon Zernike basis states as 
\begin{align}
    |z_{n_1}^{m_1}, z_{n_2}^{m_2}\rangle &= \frac{1}{\pi\sqrt{2}}\iint  \d^2q_1 \d^2q_2  \Tilde{Z}_{n_1}^{m_1}(\bq_1) \Tilde{Z}_{n_2}^{m_2}(\bq_2)
    \nn&\qquad\qquad\quad\times
    a^\dagger(\bq_1)a^\dagger(\bq_2) |0\rangle.
\end{align}
Expanding the above two-photon state in the tensor product Zernike basis, we get
\begin{align}
    |\psi_2\rangle &=     \sum_{\substack{m_1m_2\\n_1n_2}} \zeta_{n_1n_2}^{m_1m_2} |z_{n_1}^{m_1},z_{n_2}^{m_2}\rangle,\label{tpz}
\end{align}
with 
\begin{align}
    &\zeta_{n_1n_2}^{m_1m_2} \equiv \langle z_{n_1}^{m_1}, z_{n_2}^{m_2}|\psi_2\rangle
    \nn&=
    \frac{1}{\pi}\iint 
    \d^2q_1\,\d^2q_2 \, C(\bq_1,\bq_2)
    \Tilde{Z}_{n_1}^{m_1*}(\bq_1)\Tilde{Z}_{n_2}^{m_2*}(\bq_2).\label{zeta2}
\end{align}
The normalization condition \eqref{normalization2} is equivalent to 
\begin{align}
    \sum_{\substack{m_1m_2\\ n_1n_2}} |\zeta_{n_1n_2}^{m_1m_2}|^2=1.
\end{align}
\subsection{Second-order correlation function}
With the help of \eqref{field-op-1pZFrhf} and \eqref{tpz} the second-order correlation function in the Zernike basis assumes the form
\begin{align}
    G^{(2)}(\br_1,\br_2) = 4 &
    \sum_{\substack{m_1m_2m_3m_4 \\n_1n_2n_3n_4}}
    \zeta^{m_1m_2}_{n_1n_2}
    \zeta^{m_3m_4*}_{n_3n_4}\,
    \nn&\times
    \Tilde{Z}_{n_1}^{m_1} (\rho_{x_1},\rho_{y_1})\,
    \Tilde{Z}_{n_2}^{m_2} (\rho_{x_2},\rho_{y_2})
    \nn&\times
    \Tilde{Z}_{n_3}^{m_3*} (\rho_{x_1},\rho_{y_1})\,
    \Tilde{Z}_{n_4}^{m_4*} (\rho_{x_2},\rho_{y_2}).
\end{align}
\subsection{Two-photon state from the down-conversion process}
Eq. \eqref{zeta2} holds for the specific symmetric amplitude such as the one generated from the process of spontaneous parametric down conversion (SPDC) \cite{Monken1998}:
\begin{align}
    C 
    (\bq_1,\bq_2)&=\sqrt{\frac{2L}{\pi^2K}}
    v(\bq_1+\bq_2)
    \text{sinc}\left(\frac{L(|\bq_1-\bq_2|^2)}{4K}\right)
    \nn&\propto v(\bq_1+\bq_2),\quad \text{in thin crystal approx. }\label{thin crystal.}
\end{align}
Here $v(\bq_p)$ is the angular spectrum of the pump beam with $\bq_p=\bq_1+\bq_2$, $L$ is the length of the nonlinear crystal in the propagation ($z$) direction, and $K$ is the magnitude of the pump field wave vector.
The thin-crystal approximation is a well-established simplification in SPDC theory to simplify the phase-matching function when the nonlinear crystal length 
$L$ is small compared to the Rayleigh range of the pump beam. In this limit, the $sinc$ phase-matching function becomes broad in momentum space, and the two-photon amplitude is effectively determined by the spatial profile of the pump \cite{Monken1998, walborn04}.

Now, making the following change of variables
\begin{align}
    \textbf{P}=\bq_1+\bq_2,\quad
    \textbf{Q}= \frac{1}{2}(\bq_1-\bq_2),
\end{align}
\eqref{zeta2}, in the thin crystal approximation \cite{Monken1998}, takes the form 
\begin{align}
    \zeta_{n_1n_2}^{m_1m_2} &= 
    \frac{1}{\pi}
    \int \d^2\rho \, V(\bro) \, Z_{n_1}^{m_1*}(\bro)Z_{n_2}^{m_2*}(\bro),
    \label{zeta-simple1}
\end{align}
where $V(\bro)$ is the pump profile (the inverse Fourier transform of the angular spectrum of the pump). 
In the above representation, we made use of the convolution theorem, and changed the order of integration.
Expanding the pump profile in the Zernike basis $V(\bro) = \sum_{mn} a_{mn}Z_n^m(\bro)$,
\eqref{zeta-simple1} takes the form
\begin{align}
    \zeta_{n_1n_2}^{m_1m_2}&= 
    \sum_{mn}a_{mn} A_{n_1n_2n}^{m_1m_2m},\label{zetfin}
\end{align}
where we used the orthogonality of Zernike polynomials and the relation \eqref{zprod} \cite{vanhaver}.
The property $m=m_1+m_2$ of the Clebsch-Gordan coefficients guarantees that the azimuthal frequency of the initial aberration is conserved in the down-conversion process.
In principle, this conservation could be satisfied by fields that exhibit a classical or quantum correlation (entanglement) of aberrations. 
The other conditions on the Clebsch-Gordan coefficients lead to additional conservation criteria. We note here that the conditions are not restrictions on individual indices of signal and idler photons but involve sums and differences. 
Thus, there is, indeed, a correlation between the indices of the signal and the idler fields.
In the next subsection, we show that these conservation conditions lead to aberration entanglement of the down-converted photons.\

\subsection{Entanglement in Zernike modes}\label{Zentanglement}
We consider the pump profile 
having a single Zernike mode $Z_n^m$, so that the expansion coefficients \eqref{zetfin} are given only by $a_{mn}A_{n_1n_2n}^{m_1m_2m}$.

\begin{figure}[htbp]
    \centering
    \includegraphics[width=0.9\linewidth]{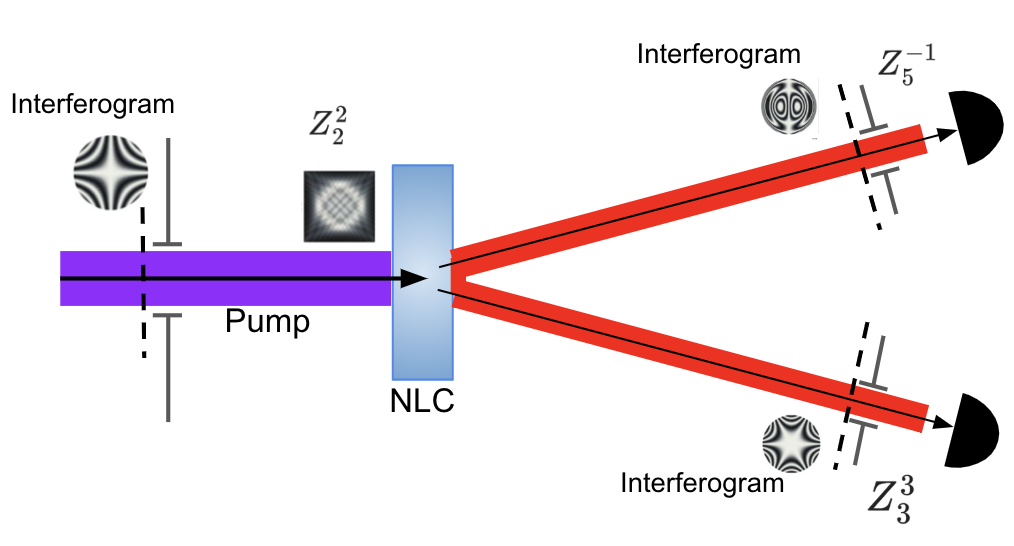}
    \caption{Schematic of Two-Photon Zernike mode entanglement via SPDC. A pump beam prepared in a Zernike mode $Z_n^m$ illuminates a thin nonlinear crystal, producing signal and idler photons whose spatial modes are correlated. The dashed lines in front of the pupils represent the interferograms pictured near them. The picture near the non-linear crystal (NLC) represents the propagated pump profile according to Eq. \eqref{ZFraunh}. The conservation of azimuthal quantum number $(m = m_1 + m_2)$ and triangle conditions for $n-$ indices, arising from Clebsch–Gordan coefficients, govern the resulting two-photon entangled state in the Zernike basis.}
    \label{fig:setup}
\end{figure}
Now, the reduced density matrix for, say, the signal photon of the two-photon
state \eqref{tpz} is given by
\begin{align}
    \rho_1 &= 
    \sum_{n_1m_1}\sum_{n_1'm_1'} \Xi_{n_1n_1'}^{m_1m_1'} \big|z_{n_1}^{m_1}\rangle\langle z_{n_1'}^{m_1'}\big|,
\end{align}
where $\Xi_{n_1n_1'}^{m_1m_1'}\equiv \sum_{n_2m_2}\zeta_{n_1n_2}^{m_1m_2} \zeta_{n_1'n_2}^{m_1'm_2*}.$
Note that 
\begin{align}
    \Xi_{n_1n_1'}^{m_1m_1'} = \Xi_{n_1'n_1}^{m_1'm_1*}.\label{symm}
\end{align}
If the above density matrix represents a mixed state then $|\psi_2\rangle$ is entangled.
From the properties of density matrices $\text{Tr}\rho_1=1$ and $\rho_1\geq 0$  
it follows that
\begin{align}
    \sum_{n_1m_1}\Xi_{n_1n_1}^{m_1m_1} =1, 
    \quad \Xi_{n_1n_1}^{m_1m_1}  \geq 0. \label{posdef}
\end{align}
The condition $\text{Tr}\rho^2\leq 1$ is equivalent to $\sum_{\substack{n_1m_1\\n_1'm_1'}}
    \left|\Xi_{n_1n_1'}^{m_1m_1'}\right|^2
    \leq 1,$
where we used the orthogonality of the Zernike modes and the symmetry property \eqref{symm}. Subtracting $\sum_{n_1m_1}\Xi_{n_1n_1}^{m_1m_1}
\sum_{n_1'm_1'}\Xi_{n_1'n_1'}^{m_1'm_1'}=1$ (cf.\eqref{posdef}) from each side of the last inequality, we have
\begin{align}
\sum_{\substack{n_1m_1\\ n_1'm_1'}}
    \left[ \left|\Xi_{n_1n_1'}^{m_1m_1'}\right|^2 - \Xi_{n_1n_1}^{m_1m_1} \Xi_{n_1'n_1'}^{m_1'm_1'} \right] \leq 0.\label{ksb}
\end{align}
Since $\rho_1$ is a positive semidefinite operator, its elements $\Xi_{n_1n_1'}^{m_1m_1'}=\langle z_{n_1}^{m_1}|\rho_1|z_{n_1'}^{m_1'}\rangle$ obey the generalized Cauchy-Schwartz-Buniakowski inequality,
\begin{align}
    \left| \Xi_{n_1n_1'}^{m_1m_1'} \right|^2 \leq \Xi_{n_1n_1}^{m_1m_1} \Xi_{n_1'n_1'}^{m_1'm_1'},
\end{align}
for any $n_1, m_1, n_1', m_1'$, which means that all the terms in the summation in equation \eqref{ksb} are either zero or negative. Then, if at least one $\left|\Xi_{n_1n_1'}^{m_1m_1'}\right|^2 - 
\Xi_{n_1n_1}^{m_1m_1}\Xi_{n_1'n_1'}^{m_1'm_1'}$ is nonzero,
the \textit{equality} in \eqref{ksb} cannot hold, indicating that $\rho_1$ is a mixed state ($\text{Tr}[\rho^2_{1}]<1$)
and, consequently, $|\psi_2\rangle$ must be entangled. 
Note that for $\left|\Xi_{n_1n_1'}^{m_1m_1'}\right|^2 - \Xi_{n_1n_1}^{m_1m_1} \Xi_{n_1'n_1'}^{m_1'm_1'}<0$ it must be that $\Xi_{n_1n_1}^{m_1m_1} >0$ and $\Xi_{n_1'n_1'}^{m_1'm_1'}>0$. 
With the help of \eqref{CG0} we see that
\begin{align}
    \text{parity}(m_1+m_2)&=\text{parity}(m),\\ \text{parity}(m_1'+m_2)&=\text{parity}(m),
\end{align}
hence it must be that
\begin{align}
    \text{parity}(m_1)&=\text{parity}(m_1'),
\end{align}
otherwise $\Xi_{n_1n_1'}^{m_1m_1'}=0$. Thus, even if parity($m_1$)$\neq$ parity($m_1'$), there still can be terms with $\Xi_{n_1n_1}^{m_1m_1} \neq 0$ and $\Xi_{n_1n_1}^{m_1'm_1'} \neq 0$ but $\Xi_{n_1n_1'}^{m_1m_1'}=0$. Noting also the second condition \eqref{posdef}, we conclude that $|\psi_2\rangle$ is entangled in the Zernike modes.

\section{Discussion}
The presence of entanglement in the Zernike basis is not merely a formal consequence of the mathematical framework — it offers practical utility for quantum optics and quantum information science. 
Like Laguerre-Gaussian or Hermite-Gaussian modes, which are the eigenmodes of free-space propagation, Zernike modes directly correspond to optical aberrations and are the natural eigenmodes of wavefront distortions. This makes them particularly appealing for quantum tasks involving real optical systems, where aberrations naturally occur and must be measured, corrected, or exploited.

One key application is quantum aberration sensing. In this protocol, a pair of photons entangled in Zernike modes is generated via spontaneous parametric down-conversion (SPDC). One photon (the probe) is transmitted through an unknown or distorted optical element, while the other (the reference) is routed directly to a detection system. By performing joint measurements in the Zernike basis, one can extract information about the aberration-induced mode transformation of the probe photon. Since Zernike modes diagonalize common aberration types (e.g., defocus, astigmatism, coma), the structure of the entangled state enables efficient inference of aberrations with reduced measurement complexity compared to more generic mode decompositions.

Another promising direction is in high-dimensional quantum communication over aberrated channels. In many realistic scenarios, including propagation through the atmosphere, tissue, or imperfect lenses, channel distortions can be compactly described in terms of Zernike polynomials. In this setting, pre-compensation or post-correction of Zernike-mode entangled photons could enable more stable transmission of qudits or entangled states through realistic, imperfect optical media.

\section{Conclusion} \label{conclusion}
We have developed a complete theoretical framework for using Zernike modes as a structured, physically grounded basis for quantum information processing. Leveraging the orthogonality, rotational symmetry, and completeness of Zernike polynomials on the unit disc, we constructed single- and two-photon wave packets and showed how these modes evolve under free-space propagation and optical transformations. We demonstrated that two-photon states generated via SPDC are entangled in the Zernike basis, with correlations arising from angular momentum conservation and Clebsch–Gordan selection rules. The results suggest that Zernike modes can serve as a new degree of freedom for quantum information tasks, particularly in systems affected by optical aberrations. We hope that this work encourages further experimental exploration of Zernike-mode-based quantum optics and motivates new applications at the intersection of spatial mode control and quantum technologies.

\textbf{Funding.} This work was supported by SCS of Armenia, grant No. 24FP-1F030. 

\textbf{Acknowledgments.}
The authors thank A.Allahverdyan, V.Mkrtchian and V.Abgaryan for discussions. 

\bibliography{references}

\end{document}